\newif\ifanonymousversion
\anonymousversionfalse

\ifanonymousversion

\documentclass[conference, compsoc]{IEEEtran}

\else

\documentclass[conference, compsoc]{IEEEtran}

\fi

\usepackage{tikz}
\usepackage{amsmath}
\usepackage{xcolor}
\usepackage{filecontents}

\usepackage{hyperref}
\usepackage{soul}
\usepackage[braket, qm]{qcircuit}
\usepackage{physics}
\usepackage{enumitem}
\usepackage{multirow}
\usepackage{booktabs}
\usepackage{algorithmicx}
\usepackage{algpseudocode}
\usepackage{algorithm}

\usepackage{caption}
\usepackage{subcaption}

\usepackage{balance}
\usepackage{nicematrix}

\usepackage{subcaption}

\usepackage{amsfonts}
\usepackage{flushend}
\usepackage{url}

\newcommand{\hbarhorizline}{\mathchar'26\mkern-7mu h}

\newcommand{\nsf}[1]{\href{https://www.nsf.gov/awardsearch/showAward?AWD_ID=#1}{#1}}

\newcommand\name[1][]{pulsejacking}
\newcommand\Name[1][]{Pulsejacking}

\begin{document}

\date{}

\title{Security Attacks Abusing Pulse-level Quantum Circuits}

\ifanonymousversion

\author{Anonymous}

\else

\author{
\IEEEauthorblockN{Chuanqi Xu}
\IEEEauthorblockA{\textit{Dept. of Electrical \& Computer Engineering} \\
\textit{Yale University}\\
New Haven, CT, USA \\
chuanqi.xu@yale.edu}
\and
\IEEEauthorblockN{Jakub Szefer}
\IEEEauthorblockA{\textit{Dept. of Electrical \& Computer Engineering} \\
\textit{Yale University}\\
New Haven, CT, USA \\
jakub.szefer@yale.edu}
}

\fi

\maketitle

\begin{abstract}
This work presents the first thorough exploration of the attacks on the interface between gate-level and pulse-level quantum circuits and pulse-level quantum circuits themselves. Typically, quantum circuits and programs that execute on quantum computers, are defined using gate-level primitives. However, to improve the expressivity of quantum circuits and to allow better optimization, pulse-level circuits are now often used. The attacks presented in this work leverage the inconsistency between the gate-level description of the custom gate, and the actual, low-level pulse implementation of this gate. By manipulating the custom gate specification, this work proposes numerous attacks: qubit plunder, qubit block, qubit reorder, timing mismatch, frequency mismatch, phase mismatch, and waveform mismatch. This work demonstrates these attacks on the real quantum computer and simulator, and shows that most current software development kits are vulnerable to these new types of attacks. In the end, this work proposes a defense framework. The exploration of security and privacy issues of the rising pulse-level quantum circuits provides insight into the future development of secure quantum software development kits and quantum computer systems.

\end{abstract}

\section{Introduction}
\label{sec:introduction}

Quantum computing is advancing rapidly with more and bigger quantum computers coming online every year: from quantum computers with one or two qubits two decades ago, to computers with 1121 qubits presently available; and projections for quantum computers with 200-qubit error corrected modularity capable of running 100 million gates by the current decade's end~\cite{ibmqubit}. These existing quantum computers are categorized as Noisy Intermediate-Scale Quantum (NISQ) devices~\cite{preskill2018quantum} since they do not provide error correction~\cite{Devitt_2013} and have noisy qubits and operations. Nevertheless, they show promise in applications like optimization, natural sciences, artificial intelligence, finance, etc~\cite{lanyon2010towards,jones1998implementation,mermin2007quantum, biamonte2017quantum, farhi2014quantum, orus2019quantum, herman2023quantum}. 

Today, quantum computers are easily accessible through cloud-based services such as IBM Quantum~\cite{ibm_quantum}, Amazon Braket~\cite{braket}, or Microsoft Azure~\cite{azure}. These services are not just for academics, but actual pay-as-you-go cloud services that anybody can use. Many startups and companies do not have their own quantum computers, and they depend on cloud-based quantum computers to run their, often proprietary, quantum circuits. For example, J.P. Morgan Chase leverages cloud-based Quantinuum H-series to develop algorithms for solving linear systems on quantum
hardware~\cite{yalovetzky2024solving}.
As the utilization of quantum computers is on the rise, the importance of securing quantum circuits and programs that execute on the quantum computers, also becomes increasingly evident. 

Quantum circuits serve as the foundational building blocks for quantum computation, orchestrating the manipulation and processing of quantum bits. These circuits encode quantum algorithms and enable the execution of complex computations, offering the potential to outperform classical computers in specific tasks. Typical quantum circuits are usually built using quantum gates, which are abstract units that define specific operations on qubits.

However, quantum circuits are not restricted to be defined only using quantum gates. 
In fact, quantum gates are only one abstract layer, and they need to be compiled into actual operations on qubits by quantum software development kits (SDKs). For example, quantum gates on superconducting quantum computers are realized by electromagnetic pulses. Recent advancements highlight the substantial advantages of deconstructing gate-level circuits into these lower layers.
Defining circuits using low-level control pulses proves advantageous for tasks such as quantum circuit decomposition, compilation, and optimization~\cite{10.1145/3307650.3322253, 10.1145/3406309, 9345604, 9797403, 9951219, 10247923}, as well as enhancing the expressivity and efficiency of quantum circuits~\cite{9951187, liang2024napa, liang2023advantages, liang2023spacepulse}. Despite the many benefits of specifying quantum circuits using lower-level control pulses, security and privacy have seldom been touched on. One recent research work introduced higher-energy state attacks, exploiting control pulses to excite qubits into higher-energy levels, and then abusing the properties of higher-energy states to circumvent standard operations in quantum computers~\cite{10.1145/3576915.3623104}.

To address research gaps in this field, this paper undertakes the first comprehensive exploration of security and privacy aspects concerning quantum circuits built partly or wholly from low-level pulses rather than purely from gates. In particular, we introduce a new class of attacks, which leverage the inconsistency between the definition of the custom quantum gate, and the low-level implementation of the pulses that realize the gate operations. For the attacks, we assume attackers can clandestinely implant or modify the low-level pulse implementation such that it is not revealed to the high-level gate definition. The proposed attacks are particularly dangerous for two reasons: Firstly, most current quantum software development kits are deficient in the interface between gate-level and pulse-level, and pulse configurations specifying pulse-level gates are in the analog form. These make our attacks less understandable and difficult to verify. Secondly, pulse configuration is determined by the hardware properties and environment. Due to reasons such as variations in manufacturing and the continuously changing environment within and near the quantum computer, pulse configuration is distinct and unique for a specific quantum hardware and needs frequent updates to maintain proper functionality. This, in turn, means that pulse specifications are individualized and time-restricted. It is almost impossible to verify pulses by setting an overall standard for pulses for all quantum computers, or by comparing them to some prior, correct pulses, since even correct pulses will have changed with time. The pulse data is updated as often as the quantum computers are calibrated, which can be every day, or even every few hours, providing space for adversaries to install attacks. Both factors result in a unique new attack surface in the quantum software supply chain, which is unique and different from classical computing, making the attacks challenging to detect and prevent.

\begin{figure}[t]
     \centering
     \includegraphics[width=0.47\textwidth]{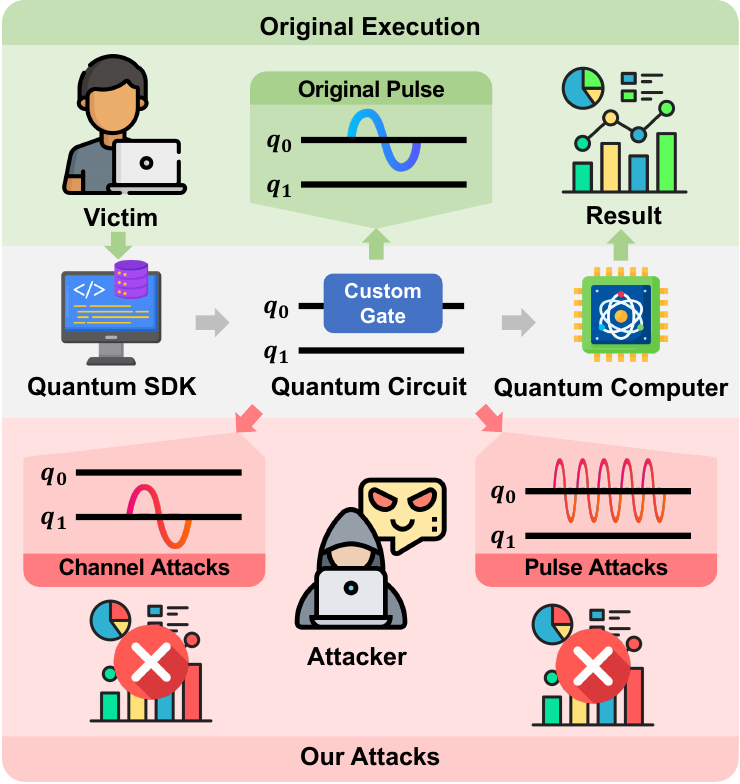}
     \caption{\small The workflow of a quantum computer, demonstrating where the channel and pulse attacks proposed in this paper can occur. The quantum software development kit (Quantum SDK) is used to develop Quantum Circuits which are then executed on a Quantum Computer. The proposed attacks target the Quantum Circuit specification and can be deployed, among others, through supply chain attacks.}
    \label{fig:overview}
\end{figure}

The high-level workflow schematic of our proposed attacks is shown in Figure~\ref{fig:overview}. Our attacks can primarily manifest themselves in two places in the quantum software development stack. The first location is the interface between gate-level and pulse-level circuits, referred to as \textit{channel attacks}. These attacks, encompassing \textit{qubit spoilage}, \textit{qubit reorder}, and \textit{qubit block}, focus more on software development kit aspects. We scrutinized two widely used quantum SDKs: Qiskit and Amazon Braket Python SDK, and identified the possibility of exploiting \textit{channel attacks} in both SDKs. The second location is in the pulse configuration, which directly specifies the pulse details. We categorize this as \textit{pulse attacks}. These attacks include \textit{timing mismatch}, \textit{frequency mismatch}, \textit{phase mismatch}, and \textit{waveform mismatch}, each exploiting specific features of pulses to manipulate qubits. We demonstrate these attacks on the real quantum computer and simulator, focusing on extensively utilized quantum algorithms: quantum teleportation, Grover's search, and quantum neural networks.

To counteract these attacks, we present a defense framework that checks each type of the discussed attacks, as well as propose a standard for the reuse of pulse-level circuits. Our defense strategy encompasses a multi-faceted approach, leveraging both hardware and software components to fortify quantum circuits against the attacks.

\section{Background}
\label{sec:background}

This section aims to offer background on how data is manipulated in quantum computers and a typical workflow for running quantum computing programs.

\subsection{Qubits and Quantum States}

In the realm of quantum computing, the fundamental unit of information is the quantum bit, or qubit. Conceptually akin to the classical bit in conventional computing, a qubit possesses two basis states, symbolized in bra-ket notation as $\ket{0}$ and $\ket{1}$. However, unlike classical bits which are constrained to values of either 0 or 1, a qubit can exist in a superposition of these states: $\ket{\psi} = \alpha \ket{0} + \beta \ket{1}$, where $\alpha, \beta$ are complex and $|\alpha|^2 + |\beta|^2 = 1$.

Qubits are often represented using vector representation. For instance, the single-qubit basis states can be denoted as two-dimensional column vectors: $\ket{0} = \begin{bmatrix} 1, \ 0 \end{bmatrix}^T$ and $\ket{1} = \begin{bmatrix} 0, \ 1 \end{bmatrix}^T$, where $T$ is the transpose. Thus, the state $\ket{\psi}$ can be expressed as $\ket{\psi} = \alpha \ket{0} + \beta \ket{1} = \begin{bmatrix} \alpha, \ \beta \end{bmatrix}^T$. Extending this notion to $n$ qubits, the space of $n$-qubit states encompasses $2^n$ basis states, ranging from $\ket{0\dots 0}$ to $\ket{1\dots 1}$. Consequently, an $n$-qubit state $\ket{\phi}$ can be represented as: $\ket{\phi} = \sum_{i = 0}^{2^n - 1} a_i \ket{i}$, where $\sum_{i = 0}^{2^n - 1}|a_i|^2 = 1$.

\subsection{Quantum Gates}

In the domain of quantum computing, fundamental operations are represented as quantum gates. The quantum circuits and programs execute a sequence of quantum gates to manipulate qubits towards desired states.

A quantum gate $U$ must be unitary, i.e., $U U^\dagger = U^\dagger U = I$, where $U^\dagger$ signifies the conjugate transpose of $U$, and $I$ is the identity matrix. Operating on a qubit $\ket{\psi}$, a quantum gate $U$ transforms it as $\ket{\psi} \rightarrow U \ket{\psi}$. Employing the matrix representation, $n$-qubit quantum gates are expressed as $2^n \times 2^n$ matrices. Quantum gates can be classified as single-qubit gates or multi-qubit gates. U3 gate is one general single-qubit gate that is parameterized by 3 Euler angles. Some other single-qubit gates include Pauli-$X$ gate: a single-qubit gate similar to the classical NOT gate, flips $\ket{0}$ to $\ket{1}$ and vice versa; {\tt RZ} gate: a phase shift between $\ket{0}$ and $\ket{1}$. Unlike classical computing, multi-qubit gates can create quantum entanglement. One notable example is the CNOT gate, also known as the CX gate, a two-qubit gate that applies a Pauli-$X$ gate to the target qubit if the control qubit is in state $\ket{1}$, otherwise, it remains unchanged. Another example is the control U3 gate, which also uses one bit to control the application of the U3 gate.

Matrix representations of some gates are presented below, adhering to the qubit order specified by Qiskit~\cite{Qiskit}:

\begin{equation*}
{\tt U3(\theta, \phi, \lambda)}=\begin{bmatrix}
\cos(\frac{\theta}{2}) & -e^{i\lambda} \sin(\frac{\theta}{2}) \\
e^{i\phi} \sin(\frac{\theta}{2}) & e^{i(\phi + \lambda)} \cos(\frac{\theta}{2})
\end{bmatrix}
\end{equation*}
\begin{equation*}
{\tt RZ(\theta)}=\begin{bmatrix}
e^{-i\frac{\theta}{2}} & 0 \\
0 & e^{i\frac{\theta}{2}}
\end{bmatrix},
{\tt CX} = \begin{bmatrix}
1 & 0 & 0 & 0 \\
0 & 0 & 0 & 1 \\
0 & 0 & 1 & 0 \\
0 & 1 & 0 & 0
\end{bmatrix}
\end{equation*}

Research proved that any unitary quantum gate can be approximated with minor error using a small set of quantum gates~\cite{deutsch1995universality}. Consequently, contemporary quantum computers typically support a limited set of basis gates. By amalgamating these basis gates, they can emulate other quantum gates. These fundamental gates, termed native gates, serve as pivotal configurations for quantum processors.

\subsection{Control Pulses}
\label{sec::superconducting_quantum_computer_controls}

Microwave pulses usually serve as the primary means to control qubits. A control pulse is analog and specified by a set of parameters, such as the envelope, frequency, and phase. The envelope delineates the shape of the signal generated by the arbitrary waveform generator (AWG), a standard laboratory instrument; frequency specifies a periodic signal utilized to modulate the envelope signal; phase manipulates the frame phase of the equipment. Together, these components constitute the output signal directed towards the qubit.

Envelopes are often discretized into time steps, with each value denoting the amplitude at a specific time step. An alternative approach involves parameterized pulses determined by predefined shapes. These parameters typically encompass the duration (pulse length), amplitude (relative pulse strength), and other factors dictating pulse shape.

Pulses for all native gates on current quantum computers are often predefined, but their parameters are updated periodically. The parameters of pulses are determined by many factors such as qubit properties and environment, and thus they are not fixed over time and are unique for a specific quantum computer: Due to the constantly changing environment, the physical properties of quantum computers also continuously vary. As a result, the parameters of pulses are regularly updated over time via automated measurements and calibrations to maintain high fidelity. Because of this, gate-level circuits need to be recompiled before they are executed so that the most recent pulses representing the gates are used. Apart from changing continuously, due to reasons such as variations in manufacture and environment, the parameters of pulses are also distinct for qubits. Parameters of pulses for different qubits on the same quantum computer, qubits on different quantum computers with the same type and model, have different parameters.

\subsection{Pulse-Level Quantum Circuit}

Quantum circuits described by quantum gates are called gate-level circuits. The gate-level circuits are transformed into pulse-level circuits by the compiler so that they can be executed on the quantum computer, or the user can directly write pulse-level circuits themselves. To realize quantum circuits, the control pulses must be properly arranged, such as the starting time of the pulses, and to which channels pulses should be applied. All of this information constitutes the pulse-level quantum circuit. Pulse-level circuits hold significant value for users. For example, pulse-level circuits are actively investigated for optimization and applications as mentioned in Section~\ref{sec:introduction}, improving both the expressivity and efficiency. 

How to incorporate pulses into quantum circuits is the design choice of quantum software development kits. One widely used method is to define the custom gates, and specify what effect these custom gates should have on qubits. Functioning similarly to normal quantum gates, custom gates require the specification of the number of qubits and the target qubits for application within a quantum circuit. The detailed operations of custom gates are either independently specified as a list of quantum gates or linked to specifications of pulse-level controls. Custom gates can be recursively defined to encompass another custom gate.

Pulse-level circuits are necessary because they specify the actual operations on quantum computers. While they are not public on all quantum computers due to complexity, an increase in exposing them has been seen, including IBM Qiskit~\cite{qiskit_pulse}, Amazon Braket~\cite{amazon_braket_pulse}, PennyLane~\cite{pennylane_pulse}, etc. Though the advantages of pulse-level circuits are widely explored, security and privacy aspects have not been researched, which are the focus of this paper.

\subsection{From Gate-Level to Pulse-Level}
\label{sec:running_quantum}

Typically, quantum circuits and programs are defined using gate-level primitives, with quantum software development kits such as Qiskit~\cite{Qiskit}, Amazon Braket Python SDK~\cite{braket}, Q\#~\cite{azure}, Cirq~\cite{cirq_developers_2022}.
Similar to compilation in classical computing, gate-level quantum circuits often need to be modified so that they satisfy the requirements of specific hardware.
This process involves decomposing non-native quantum gates into native gates, optimizing gates by grouping and eliminating redundant gates, mapping logical qubits to physical qubits, routing circuits within the constrained topology, etc.

The next step is to transform gate-level circuits to lower-level specifications, usually pulse-level circuits. This process is like assembling in classical computing. Leveraging pre-calibrated data for native gates on individual qubits or qubit pairs, microwave pulse sequences are generated and tailored for specific hardware. Besides native gates, arbitrary pulses can also be included with custom gates. Consequently, pulse-level circuits can be processed by quantum computer controllers to generate the actual operations manipulating qubits on the target quantum computer.

\section{Threat Model}
\label{sec:threat_model}

\begin{table*}[]
\centering
\begin{tabular}{|c||cccc||cc|}
\hline
\multirow{2}{*}{\textbf{Victim Level}} & \multicolumn{4}{c||}{\textbf{Victim Verification Capability}}                                                                                                                                                                                                                                                    & \multicolumn{2}{c|}{\textbf{Vulnerability to Our Attacks}}      \\ \cline{2-5}  \cline{6-7} 
                                       & \multicolumn{1}{c|}{\textbf{Gate-Level}} & \multicolumn{1}{c|}{\textbf{\begin{tabular}[c]{@{}c@{}}Gate-Pulse\\ Interface\end{tabular}}} & \multicolumn{1}{c|}{\textbf{\begin{tabular}[c]{@{}c@{}}Pulse-Level\\ Syntax\end{tabular}}} & \textbf{\begin{tabular}[c]{@{}c@{}}Pulse-Level\\ Semantics\end{tabular}} & \multicolumn{1}{c|}{\textbf{Channel Attack}} & \textbf{Pulse Attack} \\ \hline \hline
\textbf{Level 1}                       & \multicolumn{1}{c|}{}                    & \multicolumn{1}{c|}{}                                                                        & \multicolumn{1}{c|}{}                                                                      &                                                                          & \multicolumn{1}{c|}{Can Attack}            & Can Attack          \\ \hline
\textbf{Level 2}                       & \multicolumn{1}{c|}{$\checkmark$}        & \multicolumn{1}{c|}{}                                                                        & \multicolumn{1}{c|}{}                                                                      &                                                                          & \multicolumn{1}{c|}{Can Attack}            & Can Attack          \\ \hline
\textbf{Level 3}                       & \multicolumn{1}{c|}{$\checkmark$}        & \multicolumn{1}{c|}{$\checkmark$}                                                            & \multicolumn{1}{c|}{}                                                                      &                                                                          & \multicolumn{1}{c|}{---}                        & Can Attack          \\ \hline
\textbf{Level 4}                       & \multicolumn{1}{c|}{$\checkmark$}        & \multicolumn{1}{c|}{$\checkmark$}                                                            & \multicolumn{1}{c|}{$\checkmark$}                                                          &                                                                          & \multicolumn{1}{c|}{---}                        & Algorithm-Dependent   \\ \hline
\textbf{Level 5}                       & \multicolumn{1}{c|}{$\checkmark$}        & \multicolumn{1}{c|}{$\checkmark$}                                                            & \multicolumn{1}{c|}{$\checkmark$}                                                          & $\checkmark$                                                             & \multicolumn{1}{c|}{---}                        & ---                      \\ \hline
\end{tabular}
\caption{Victim classification based on victim's capabilities to verify gate-level and pulse-level circuits, and attacks that can be effectively applied. Checkmark means the victim has the corresponding capability to verify the code at that level.}
\label{tab:victim_classification}
\end{table*}

In this paper, we introduce attacks targeting quantum circuits that incorporate pulse-level controls. Typically, pulse-level controls are integrated into gate-level circuits using user-defined custom gates.

\subsection{Assumptions}

The target of the attacks is quantum circuits that contain custom pulse-level controls. We assume that attackers are able to modify the specification of the control pulses on victims' quantum circuits. We assume that after the malicious modifications have been performed unknowingly to victims, victims execute the circuit, thus triggering the operation of the modified pulse gates.

Pulse attacks can be performed in two distinct ways. The first way involves the supply chain attack, where attackers are assumed to perform malicious modifications at various points in the software development process. This could include directly altering local code, tampering with public repositories from which victims download code, or using malicious toolchains, such as compromised compilers, to change pulse information covertly. Importantly, the supply chain attack only requires attackers to be able to modify user-level programs secretly. It does not depend on the ability to alter downstream components in the workflow, nor the ability to interfere with the infrastructure that executes quantum programs. This type of attack will be the main focus of this paper.

The second way involves more powerful attackers who are capable of altering the infrastructure itself. Instead of manipulating user-level quantum programs, these attackers can modify the infrastructure so that the pulse specifications do not produce the expected results. For instance, an attacker might manipulate the environment so that the pulse specifications in the program no longer accurately correspond to the actual conditions in the hardware. This will be left for future research.

While it is conceivable for attackers to modify other parts of quantum circuits, and this could be combined with our attacks, we assume attackers only alter pulse configurations for quantum circuits. This assumption is rooted in the clandestine nature of the attack. We argue that detecting the installation of the attack becomes more challenging for victims when only pulse configurations are modified, as opposed to modifications on other parts. For instance, gate-level modifications can be easily detected through methods performing exact match checks of previous and current quantum circuits, such as quantum computer antivirus~\cite{9840181, 10133711} or maybe simply the hash check. Conversely, modifications on pulse configurations remain more secret. To the best of our knowledge, no current tool exists that could verify pulse-level quantum circuits, and our work raises the necessity of verifying quantum circuits at the lower level.

\subsection{Attack Stealthiness}
\label{sec:attack_stealthiness}

All victims are assumed to understand gate-level circuits and any higher levels above gate-level circuits, but their capabilities are distinguished by how much they can verify the pulse-level circuits. The verification of pulse-level circuits requires the two-fold explainability of pulses: 
\begin{enumerate}
    \item Syntax: relations between pulses and logic operations, more specifically, the effect of pulses on qubits. E.g., the pulse rotates a qubit 90 degrees along the X axis.
    \item Semantics: relations between pulses and meanings or motivations. E.g., applying a Hadamard gate and a CNOT gate typically creates an entanglement state.
\end{enumerate}
Syntax verification is heavily dependent on translating pulses into gates or operations, which is specific to the hardware but applicable to all quantum programs executed on that hardware. Conversely, semantics verification is contingent on the specific quantum algorithms and the underlying logic of their designs, which is general across different hardware platforms but presupposes that the pulses are syntactically correct.

The verification process at both levels poses significant challenges, thereby enhancing the stealthiness of our attacks introduced in this paper.

\subsubsection{Challenges of Syntax Verification}
\label{sec:challenges_of_syntax_verification}

The pulse-level controls are similar to embedded assembly or machine code in classical computing, enabling direct low-level hardware control. However, unlike the fixed and limited number of instruction set architectures in classical computing, pulses are defined by analog data, such as the frequency and amplitude of electromagnetic waves. We present three major challenges for the syntax verification:

\begin{itemize}[leftmargin=*]
    \item \textbf{Interpretability:} Due to analog characteristics, pulse data is not easily interpretable by humans. Moreover, since this data is hardware-dependent, different qubits possess distinct calibration data. The same analog data does not necessarily have the same functionality on two qubits or even the same qubits under varying environmental conditions. This complexity hinders victims from understanding the actual impact of the pulses.
    \item \textbf{Volatility:} The qubit properties are frequently changing~\cite{10.1145/3600160.3600192}, and regular machine calibrations on cloud platforms accentuate the timeliness of pulse data. This means that pulse data may quickly become outdated, necessitating timely re-calibrations. Although changes in a short time frame may be minor, they can accumulate if not properly calibrated, prompting victims to seek updated pulse data. The time interval for updates depends on the hardware and properties, ranging from days to months, which is not too short to invalidate the attack quickly nor too long to free circuits from updates.
    \item \textbf{Cost:} Calibration experiments are resource-intensive and costly. In addition, victims may find these experiments complex, leading them to rely on quantum circuit providers for detailed experiment information. Even if victims conduct calibrations themselves, the intricacy of the process may still drive them to seek assistance from quantum circuit providers, and attackers may install attacks in these places rather than the pulse data directly.
\end{itemize}

\subsubsection{Challenges of Semantics Verification}
\label{sec:challenges_of_semantics_verification}

A key distinction between quantum programs and classical programs that complicates semantics verification is the nature of operations in quantum programs, which focuses on controlling the evolution of qubits. This results in control flows that are more abstract and challenging to comprehend. While this complexity is significant, it is not the only hurdle. Another critical issue is that not all quantum algorithms are inherently interpretable. For instance, parametric algorithms such as quantum deep learning or quantum approximate optimization algorithms have parameters trained to fit specific data sets. These parameters do not possess an explainable meaning, adding another layer of difficulty.

\subsection{Victim Classification}
\label{sec:victim_classification}

Based on the victim's capabilities to verify the pulse-level circuits at the syntactical and semantic level, victims are classified into 5 levels as shown in Table~\ref{tab:victim_classification}:
\begin{enumerate}
    \item Level 1: the weakest victims that cannot verify either gate-level circuits or pulse-level circuits.
    \item Level 2: victims that can only verify gate-level circuits. In Section~\ref{sec:attack}, we propose the \textit{channel attacks} which leverage the fact that the gate-level circuits are all the same but the underlying pulse-level circuits can be arbitrarily modified. Therefore, for level 1 victims, they cannot detect the channel attacks.
    \item Level 3: victims that can verify the gate-level circuits and check pulse-level circuits, but they cannot verify the pulse-level circuits at the syntactical level. Notice that victims are assumed to understand the quantum algorithms at the logic level in this case, but they still cannot verify the pulse-level circuits at the semantic level because verification at the syntactical level is the precondition. In Section~\ref{sec:attack}, we propose the \textit{pulse attacks} which modify the underlying pulse configuration.
    \item Level 4: victims that can verify the gate-level circuits and pulse-level circuits at the syntactical level, but they cannot verify the pulse-level circuits at the semantic level. The pulse attacks may apply here depending on the algorithm's interpretability.
    \item Level 5: the strongest victims that can verify the gate-level circuits and pulse-level circuits at both the syntactical level and the semantic level.
\end{enumerate}

Level 1 victims are too weak, and thus they are not the focus of this paper. On the contrary, level 5 victims are too strong. Even though they can defend themselves from the attacks proposed in this paper, we claim that the assumption of the capability to verify the pulse-level semantics for all quantum algorithms is too demanding and may not be possible, such as for quantum neural networks. Therefore, we mainly discuss the level 2 to 4 victims in this paper.

\section{Proposed Attacks}
\label{sec:attack}

\begin{figure*}[ht]
     \centering
     \includegraphics[width=0.98\textwidth]{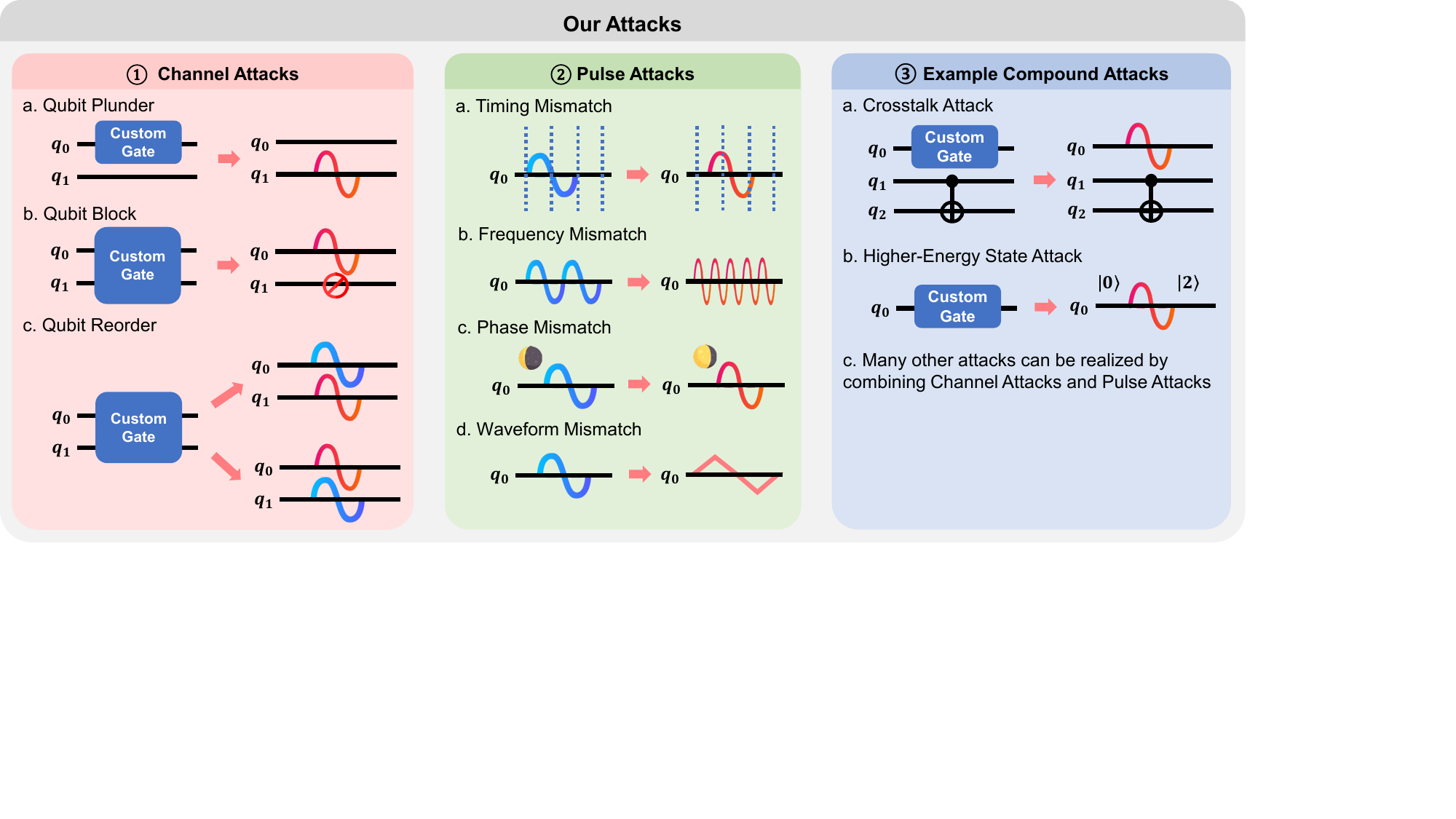}
     \caption{\small Illustration of our attacks.}
    \label{fig:attack}
\end{figure*}

In this section, we present a classification of our attacks. The pulse-level controls are typically integrated into quantum circuits through user-defined custom gates according to most current quantum software development kits. The custom gate, a gate-level operation similar to a standard quantum gate, can encapsulate pulse-level controls within its structure. Consequently, the attacks can be inserted at two primary places: the interface between the gate-level and pulse-level description, which we refer to as the \textit{channel attacks}, and the detailed configuration of the pulse-level controls, which we refer to as the \textit{pulse attacks}. Besides the distinction at the definition level, our attacks can also be performed together with themselves or with other attacks, or be abused as the approach to realize other attacks, which we refer to as the \textit{compound attacks}. The illustration of our new attacks is shown in Figure~\ref{fig:attack}.

\subsection{Channel Attacks}

The channel attacks stem from inconsistencies at the interface between the gate-level quantum gate and the pulse-level pulses. We have identified a prevalent flaw in most current quantum software development kits, where the matching between the qubits of the quantum gate and the channels of pulses is not adequately verified and sufficiently enforced. This defect leads to two types of attacks, \textit{qubit plunder} and \textit{qubit block}. In addition, the problem that the quantum gate and the pulses are considerably loosely combined entails the third type of attack, \textit{qubit reorder}:

\begin{itemize}[leftmargin=*]
    \item {\bf Qubit Plunder: } Qubit plunder arises directly from the misalignment between the qubits of the quantum gate and the channels of the underlying pulses. In most quantum software development kits, custom gates can be applied to any qubit set without corresponding checks on the channels to which the pulses are applied. For instance, a custom gate may be designated for qubits 0 and 1, while the underlying pulses could be intended for channels corresponding to qubits 2 and 3. Victims of this attack suffer when only gate-level verification is performed, as channels can be arbitrarily selected. Besides, this attack can be performed on any type of channel. With qubit plunder, a single-qubit quantum gate could maliciously affect control or readout channels, leading to unwanted entanglement or measurements. This attack may be detected when the plundered channels are already occupied with pulses, but the pulse overlap error does not clearly point to the root cause of the~problem.
    
    \item {\bf Qubit Block: } In contrast to qubit plunder, qubit block involves claiming more qubits at the gate level than the number of channels to which pulses are applied. While theoretically inconsequential in perfect quantum computers or with error correction, this attack prolongs execution time. On current NISQ (Noisy Intermediate-Scale Quantum) computers, it can induce qubit decoherence, causing unidirectional decay to $\ket 0$. For instance, one can specify an empty custom gate with only a delay inside, and attackers can control the decay rate and hence the results by manipulating the delay time of the custom gate.
    
    \item {\bf Qubit Reorder: } Apart from misalignments between gate-level qubits and pulse-level channels, the loose definition of custom gates relative to pulses allows the reordering of the encapsulated pulses. Custom gates merely serve as containers to store pulses and lack mechanisms to enforce the arrangement of pulses underneath. For instance, the direction of a custom CNOT gate can be exchanged by exchanging channels (typically also need to add or remove some pulses together). Consequently, channels associated with multi-qubit gates can be arbitrarily permuted.
\end{itemize}

The inconsistency between gate-level and pulse-level in most SDKs has not been explored before and thus can lead to security attacks. We claim the attacks stem from the inconsistency between the high-level and low-level abstractions, which appear to be problems in many fields. For instance, Spectre~\cite{8835233} and Meltdown~\cite{10.5555/3277203.3277276} in classical computers are mainly due to the inconsistency between the program instructions and microarchitecture. In comparison, for the channel attack, attackers leverage the inconsistency between program instructions (gates) and actual microarchitecture instructions (pulses).

\subsection{Pulse Attacks}

Compared with the channel attacks, the pulse attacks represent a direct manipulation of the underlying pulse data. They involve malicious modifications to various features that specify the pulses controlling the qubits. Each of these features can be exploited by attackers to execute their attacks. For a comprehensive understanding of the low-level pulse controls on qubits, readers are directed to~\cite{10.1063/1.5089550}. The attacks are characterized by the mismatch between the original pulse data and the maliciously altered pulse data:

\begin{itemize}[leftmargin=*]
    \item {\bf Timing Mismatch: } This type of attack introduces synchronization issues by altering the timing of pulses. One type of timing attack is to change the duration of the pulses and thus change the influence of the pulses, which can cause the operation to be different or also cause the decoherence similar to the qubit block. However, due to the duration of the pulse being part of the waveform of the pulses, we classify this type of attack as waveform mismatch described in the following. The precise timing of pulses and delays is crucial in the pulse definition, complicated further by constraints such as sampling time intervals, acquire alignment, pulse alignment, and granularity of the pulse. For instance, the starting time of a pulse must be the minimum common multiple of acquire alignment and pulse alignment, and the duration of a pulse must be multiples of the pulse granularity. Otherwise, the synchronization problem will cause the qubit to oscillate between $\ket 0$ and $\ket 1$. Attackers can exploit these constraints by introducing slight delays to displace pulses from the allowed increments, thereby affecting results. Fortunately, while earlier versions of Qiskit and IBM machines overlooked timing checks, subsequent updates have addressed this vulnerability. We list it here to notify this type of attack when designing the systems.

    \item {\bf Frequency Mismatch: } Quantum states $\ket 0$ and $\ket 1$ are encoded using energy levels. The frequency of a pulse must correspond to the energy difference between these energy levels. The fundamental principle in quantum mechanics dictates that only pulses with frequencies matching the energy difference between specific energy levels of the qubit can effectively control it. Therefore, attackers can manipulate the influence of a pulse by modifying its frequency. For instance, altering the frequency of a pulse to the forbidden range so that the pulse can be disabled, or to match the unwanted energy levels could drive qubits to another state, similar to higher-energy state attack~\cite{10.1145/3576915.3623104}, compromising the integrity of the quantum computation.

    \item {\bf Phase Mismatch: } In pulse-based quantum computing, the phase of a pulse is a crucial parameter that determines the rotation applied to the qubit. By altering the phase, attackers can effectively induce rotations akin to specific quantum gates, such as the rotational-Z gate commonly used for manipulating qubit states. This manipulation exploits the fact that phase changes can rapidly update the frame tracking the qubit state, allowing for instantaneous and nearly error-free rotations. Attackers may strategically introduce phase shifts at specific points in the pulse sequence to disrupt the intended quantum computation, leading to erroneous results. One advantage of phase mismatch is that it can realize any gate operation: any unitary $U \in \text{SU}(2)$ 
    can be decomposed as~\cite{PhysRevA.96.022330}:
    \begin{equation}
        U(\theta, \phi, \lambda) = Z_{\phi-\pi/2} X_{\pi/2} Z_{\pi-\theta} X_{\pi/2} Z_{\lambda-\pi/2}
    \end{equation}
    where $Z_\theta$ is the rotational-Z gate with the rotational angle $\theta$, which is the same as the phase of the pulse on many quantum computers, and $X_{\pi/2}$ is the rotational-X gate with rotational angle $\pi/2$, or {\tt SX} gate with a global phase.

    \item {\bf Waveform Mismatch: } The waveform of a pulse, representing the envelope function of its shape, influences the dynamics of qubit manipulation. A detailed understanding of pulse waveforms is essential for controlling qubit rotations accurately. A simple model mentioned in~\cite{10.1063/1.5089550} (Equation 91) gives the Hamiltonian: $\bar{H}_d = -\frac{\Omega}{2} V_0 s(t) (I\sigma_x + Q\sigma_y)$, where the $I$-component is the ``in-phase'' pulse that corresponds to rotations around the $x$-axis, and the $Q$-component is the ``out-of-phase'' pulse that corresponds to the rotations about the $y$-axis. The qubit satisfies the Schr\"{o}dinnger equation $H\ket{\psi(t)} = i \hbarhorizline \frac{\partial}{\partial t} \ket{\psi(t)}$. Simply speaking, the amplitude of the I and Q components correspond to the rotational angle along the $x$ and $y$-axis respectively. Attackers can exploit this by tampering with the waveform, thereby controlling the gates applied to qubits. For instance, by modifying the amplitude or shape of the waveform, attackers can induce unintended rotations or interfere with the coherence of qubit states.
\end{itemize}

\subsection{Compound Attacks}

The channel attacks and pulse attacks can be used as building blocks for compound attacks. In compound attacks, individual channel attacks or pulse attacks are used as a building block to achieve higher-level attack goals. We list several interesting compound attacks that can be implemented by leveraging our attacks: 
\begin{itemize}
\item \textit{Qubit Flip:} This attack aims to flip the state of a qubit, thereby altering its computational output.
\item \textit {Entanglement Injection:} Attackers may seek to inject unwanted entanglement between qubits, disrupting the coherence of the quantum system. 
\item \textit{Decoherence Amplification:} Decoherence amplification attacks aim to accelerate the decoherence process in qubits, leading to faster loss of quantum information and degradation of computational fidelity.
\item \textit{Trojan Attack:} The control qubit can be used as the trojan qubit to trigger the gates.
\item \textit{Measurement Manipulation:} manipulate the end-of-circuit or mid-circuit measurement results.
\item \textit{Multi-Tenant Attack:} entanglement can also be created in the multi-tenant environment. Though multi-tenant quantum computers are not supported yet, this paradigm is actively researched~\cite{10.1145/3310273.3321561, 9407180, Niu2023enablingmulti, upadhyay2023stealthy}, and our attacks can be applied in terms of multi-tenant use cases, especially with the channel attacks.
\end{itemize}

Moreover, our attacks can be combined with other attack vectors or leveraged as a means to facilitate other attacks. Below is an incomplete list based on related works: 
\begin{itemize}
    \item \textit{Crosstalk Attack:} Injecting pulses with significant crosstalk effects or modifying pulse frequencies to exacerbate crosstalk effect can lead to crosstalk attacks~\cite{10.1145/3370748.3406570, 9193969, 9840181, 10133711, 9251858, maurya2024understanding}, where the parallel execution of some gates leads to larger errors.
    \item \textit{Higher-Energy State Attack:} Driving qubits to a higher-energy state outside the two-level computational system $\ket 0$ and $\ket 1$, which inherently requires pulse operations, enables attackers to execute higher-energy state attacks~\cite{10.1145/3576915.3623104} that can invalidate normal quantum gates.
    \item \textit{``Horizontal" Information Leakage:} Modifying the measurement pulse can lead to ``horizontal" information leakage~\cite{10.1145/3548606.3559380, 9951250, xu2024thorough, tan2023extending} in quantum circuits that exists through executions and interferes the input states of the next execution.
    \item \textit{Power Attack:} Pulses are related to power traces of quantum circuits~\cite{10.1145/3576915.3623118, erata2024quantum}, and thus our attacks can influence the equipment power features.
    \item \textit{Fault-Injection:} Changing the pulses at the target locations can lead to fault-injection~\cite{oliveira2022qufi, xu2023classification}. This can also be done by reasoning the relations between the pulses and control electronics~\cite{das2024investigating}.
\end{itemize}
By leveraging the flexibility and intricacies of pulse-based quantum computing, attackers can orchestrate compound attacks that exploit multiple vulnerabilities in quantum systems simultaneously, posing significant challenges to quantum security protocols.

\section{Attack Demonstration}
\label{sec:demonstration}

In this section, we present attack implementations. Our attacks are general, so there are infinite implementations. We first propose one straightforward example of qubit flip. Then we demonstrate several algorithm-specific examples.

\subsection{Toy Attack Example: Qubit Flip Attack}

Given the extensive control over quantum circuits afforded by the attacks, one generic approach involves the attacker utilizing an extra layer of custom gates on all qubits either in the end to manipulate final results or at the beginning to control input states. This requires the modification of gate-level circuits, which requires attackers to be more powerful and is beyond the main threat model of this paper. This example serves to show one most straightforward way to perform the attack. Attackers may claim that the custom gate layer is for better fidelity, such as optimized readout pulses for lower measurement error or higher fidelity state preparation. Notice that level 1 victims have the lowest verification ability, and attackers can directly substitute the whole program. Here the goal is to present one simplest use case of the attack, and victims can check anything except the custom gate layer.

\begin{figure}[t]
     \centering
     \includegraphics[width=0.39\textwidth]{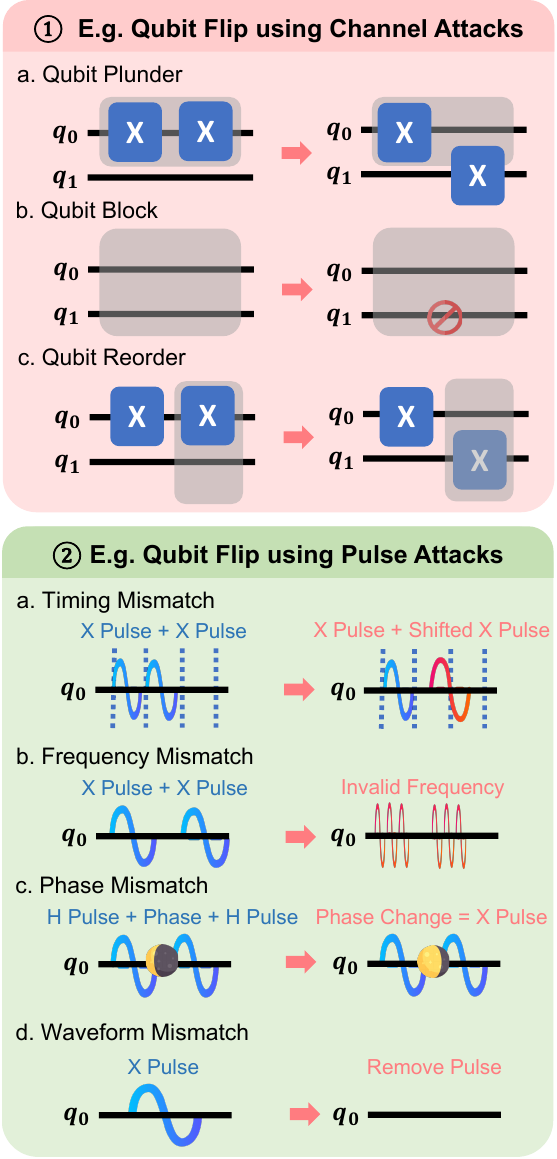}
     \caption{\small Implementations of qubit flip with channel attacks and pulse attacks. The gray block in the channel attack section represents the custom gate that is manipulated in the attacks.}
    \label{fig:generic_imp}
\end{figure}

We present some straightforward implementations of the qubit flip attack. The implementations are shown in Figure~\ref{fig:generic_imp}, but the attack clearly extends beyond mere qubit~flip.

\subsubsection{Level 1 Victim: Qubit Flip Attack using Channel Attacks}

The power of channel attacks diminishes in the following order due to their control~abilities:

\begin{itemize}[leftmargin=*]

    \item \textbf{Qubit Plunder: } If an additional channel is provided, qubit flip through qubit plunder can be realized by manipulating pulse movement. A direct configuration involves adding a layer of two Pauli-X pulses on each qubit. Two Pauli-X gates are the same as the identity gate, leaving the qubit unchanged. The qubit flip occurs when one of the Pauli-X pulses is moved to an unused channel, altering the probability between $\ket{0}$ and $\ket{1}$ on that qubit.
    
    \item {\bf Qubit Reorder: } Similarly, results can be altered by reordering pulses. A straightforward scheme entails adding two additional Pauli-X gates on one qubit, incorporating one of the Pauli-X gates in the custom gate. Thus, keeping the pulse does nothing, while reordering the pulses flips two qubits simultaneously. This is a weaker attack than the qubit plunder in that it can only flip an even number of qubits simultaneously, while qubit plunder can flip any qubit independently under the condition of extra channels.

    \item {\bf Qubit Block: } Qubit block is the least potent, incapable of arbitrary result modification. If the qubit is perfect with no decoherence or can be error-corrected, it cannot modify the results. One way as described previously is to add a long delay before the measurement, and when accounting for decoherence, qubits will decay to $\ket{0}$.
    
\end{itemize}

\subsubsection{Level 1 Victim: Qubit Flip Attack using Pulse Attacks}

Because pulse attacks can directly control the pulses, they are more powerful than channel attacks:

\begin{itemize}[leftmargin=*]

    \item {\bf Timing Mismatch: } In supported platforms, synchronization leads to qubit to oscillate between $\ket 0$ and $\ket 1$. Therefore, the results can be easily controlled by manipulating the number of units for the delay, though the ability is limited by the granularity of the machine. How the number of units for the delay corresponds to the state change can be easily measured with simple experiments.
    
    \item {\bf Frequency Mismatch:} Only frequencies correlated to energy between two energy levels can lead to energy level transitions, which means the operation of one gate can be easily disabled by tuning the frequency into the forbidden range. The scheme can be the same as qubit plunder by replacing the moving of pulse to changing of frequency, i.e., adding two Pauli-X pulses and sandwiching one frequency setting between them, and setting the frequency to the forbidden range when a qubit flip is desired.
    
    \item {\bf Phase Mismatch:} Qubit flip through phase change can be implemented with the identity $X = H \cdot RZ(\pi) \cdot H$. The custom gate is composed of 3 gates, the Hadamard gate, one Rotational-Z gate, and another Hadamard gate. When setting the phase, i.e., setting the rotational angle of the Rotational-Z gate, to be 0, the gate is identical to the identity gate. However, when setting the phase to be $\pi$, the gate is identical to the Pauli-X gate.
    
    \item {\bf Waveform Mismatch: } Waveform attack is the most flexible attack. The gate can be disabled by setting the amplitude to 0, and the rotational angle can be tuned by easily setting the waveform. The qubit flip can be easily realized by adding one empty placeholder pulse and changing the waveform of the placeholder pulse to the Pauli-X pulse when the qubit flip is wanted.
\end{itemize}

\begin{figure*}[t]
     \centering
     \begin{subfigure}[b]{0.75\textwidth}
         \centering
         \includegraphics[width=\textwidth, trim={0cm 9.3cm 2cm 0cm},clip]{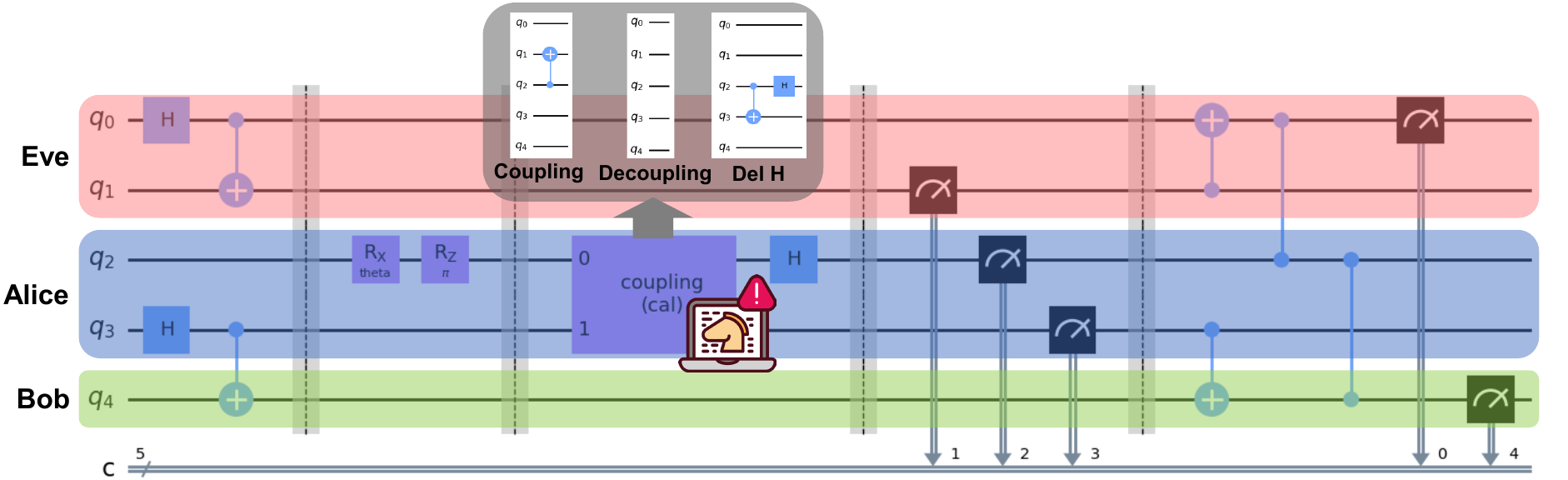}
         \caption{\small Quantum circuit of quantum teleportation. Alice (blue block) wants to transport her quantum state to Bob (green block), with the coupling gate provided by Eve (red block). Eve can tamper the pulses to achieve many goals, some of which are listed in the gray block.}
         \label{fig:qt_circ}
     \end{subfigure}~
     \begin{subfigure}[b]{0.25\textwidth}
         \centering
         \includegraphics[width=\textwidth, trim={0cm 0.3cm 0cm 0cm},clip]{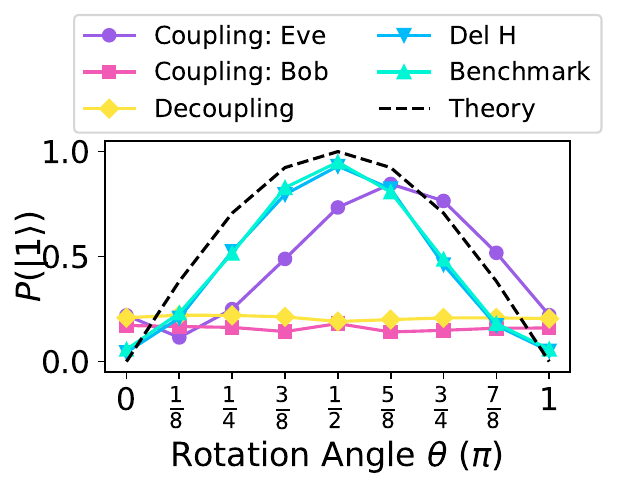}
         \caption{\small Bob's $P(\ket 1)$ (``Coupling: Eve'' is Eve's) with different pulse configurations of the coupling gate.}
         \label{fig:qt_result}
     \end{subfigure}
    \caption{\small Example demonstration of the algorithm-specific attack on quantum teleportation.}
    \label{fig:qt}
\end{figure*}

\subsection{Algorithm-Specific Attack Examples}

While the aforementioned universal implementations are simple, straightforward, and applicable to all kinds of quantum circuits, they are easily detectable through in-depth circuit examination, even by humans. For example, custom gates with only two Pauli-X pulses on each qubit before measurement usually make no sense within the quantum circuit. Though shifting the custom gates to the middle may be feasible and make detection less obvious, it typically remains implausible for arbitrarily targeted attacks. Given that the configuration of custom gates hinges on the specific quantum algorithm, such as decisions on pulse utilization and placement, a universal scheme for installing attacks to control results may not exist. Nonetheless, mapping between pulses and results may be acquired through methods such as brute force or trial and error.

A more clandestine approach involves implementing the quantum algorithm as a pulse-optimized or parametric quantum circuit, as discussed in Section~\ref{sec:introduction}. For instance, pulses employed for quantum circuit decomposition, compilation, and optimization exhibit complexity and modifiability. These pulses are typically optimized for specific quantum gates, such as Toffoli gates; or tailored for particular objectives, such as the multiply-controlled gate in Grover's search. Although these pulses serve fixed goals, they necessitate periodic updates due to changes in the physical properties of quantum computers, as explained in Section~\ref{sec:challenges_of_syntax_verification}. Another instance is pulse learning for variational quantum algorithms like quantum machine learning, where the parametric nature involves learning features of pulses during the process. These parameters, akin to weights in neural networks, are challenging to interpret individually, and may be totally different when training again depending on the initial values, making them susceptible to our attacks.

In this section, we provide some examples to demonstrate our attacks with algorithm-specific implementations. Because the stealthiness of the attack is highly dependent on the capability of the victims as discussed in Section~\ref{sec:victim_classification}, we explore the attacks on level 2 to 4 victims: channel attack on quantum teleportation for level 2 victim; pulse attack on Grover's search for level 3 victim; pulse attack on quantum neural networks for level 4 victim. The experiments for level 2 and 3 victims were done on \texttt{ibm\_osaka} on IBM Quantum, which is a 127-qubit quantum computer, while the experiments for level 3 victims were done on simulators.

\subsubsection{Level 2 Victim: Channel Attack on Quantum Teleportation}

This demonstrates the channel attack to maliciously couple and decouple qubits.

One of the most important features that differentiate quantum computing from classical computing is quantum entanglement. Qubits can be coupled together and the measurement result will depend on how they are connected, rather than their quantum states independently. Quantum entanglement is the feature that is used in almost every quantum algorithm. One example of using quantum entanglement is quantum teleportation~\cite{PhysRevLett.70.1895}, which is a technique for transferring quantum information from a sender at one location to a receiver some distance away. The quantum circuit for the demonstration is shown in Figure~\ref{fig:qt_circ}. Alice prepares her quantum state on $q_2$, and this quantum state can be transmitted to Bob through the circuit when the coupling gate in the middle is the CNOT gate. However, assuming that the coupling gate is provided as a service in the form of a custom gate with pulses that are provided by Eve, then Eve can perform many types of attacks to influence the results. For instance, we show that Eve can secretly couple or decouple qubits. 

In our demonstration, the gate-level circuit is the same for all cases, so the channel attack is hidden from the gate-level verification. Specifically, Eve can perform \textit{Coupling} $q_2$ and $q_1$ by applying the CNOT pulse even though the coupling gate is applied on $q_2$ and $q_3$, \textit{Decoupling} by simply doing nothing, or \textit{Del H} by applying the pulse of Hadamard gate to cancel Hadamard gate after the coupling gate.

In the experiment, we use a rotational-X gate with angle $\theta$ and Pauli-Z gate to prepare the state for transmission: $\ket \psi = \cos\frac{\theta}{2} \ket 0 + \sin\frac{\theta}{2} \ket 1$. The probabilities of measuring $\ket 1$ with the above pulse configurations are shown in Figure~\ref{fig:qt_result}. ``Theory'' shows the theoretical results with no errors, and ``Benchmark'' shows the results with the original quantum teleportation circuit. 

``Coupling: Eve'' shows Eve's measurement result when Eve performed the ``Coupling" gate, and it shows that Eve can secretly copy the quantum teleportation circuit to steal the quantum state in transmission, while ``Coupling: Bob'' shows that Bob cannot receive any state because the coupling of $q_2$ and $q_3$ is secretly occupied by Eve. Similarly, ``Decoupling'' does not couple $q_2$ and $q_3$, so Bob cannot receive the quantum state. Although ``Del H'' shows the same probability as the benchmark, the correct state should be a pure state $\ket \psi = \alpha \ket 0 + \beta \ket 1$, but Bob actually receives a mixed state of $\ket 0$ with probability $|\alpha|^2$ and $\ket 1$ with probability $|\beta|^2$. This attack can be successful even though we assume Bob can do simple checks on the measurement results which behave the same as the correct results.

\subsubsection{Level 3 Victim: Pulse Attack on Grover's Search}

\begin{figure}[ht]
     \centering
     \begin{subfigure}[b]{0.5\textwidth}
         \centering
         \includegraphics[width=\textwidth, trim={7cm 2cm 4cm 0.3cm},clip]{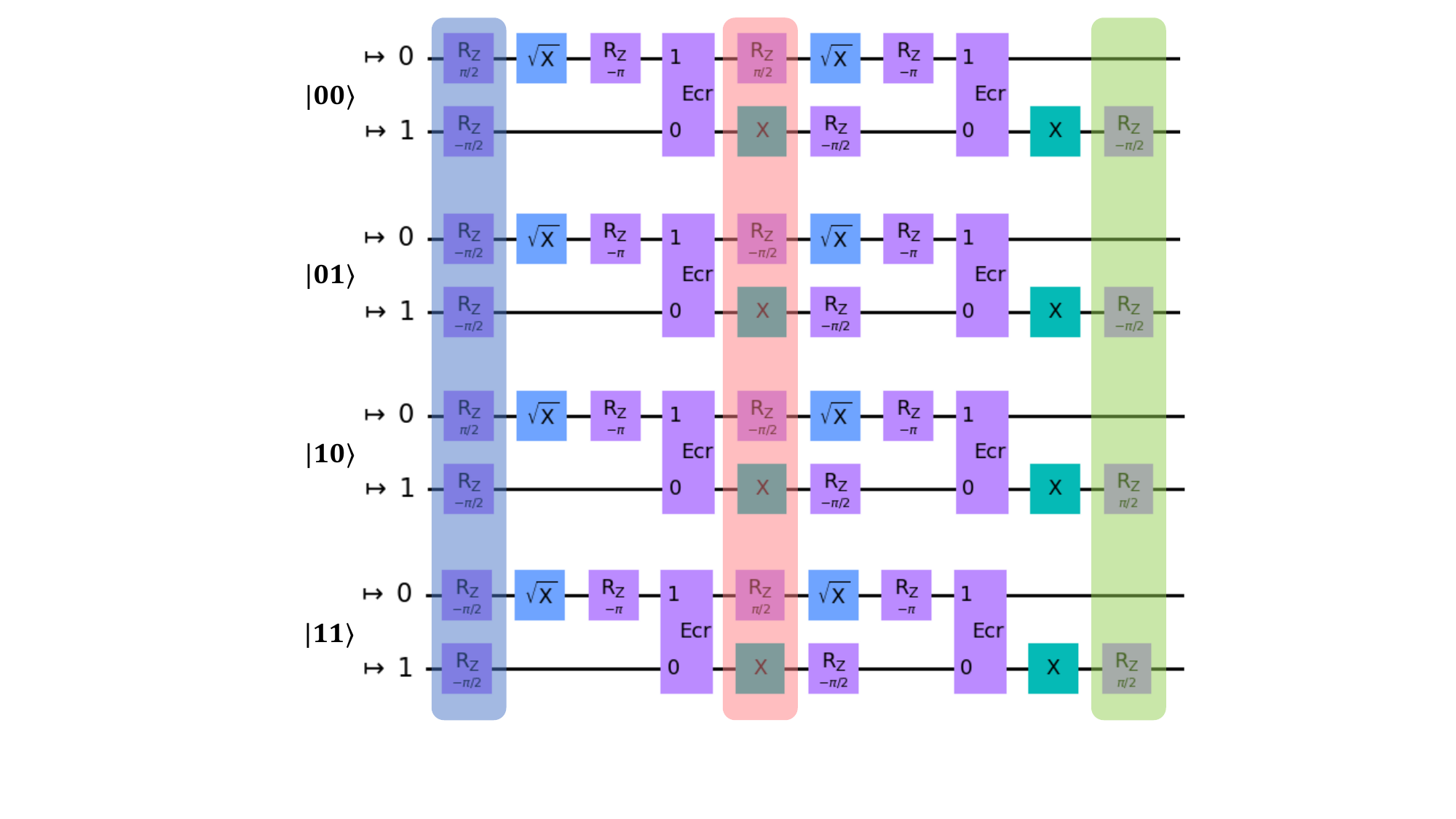}
         \caption{\small Part of the quantum circuit of 2-qubit Grover's operator. Translucent blocks show the difference for search states from $\ket{00}$ to $\ket{11}$.}
         \label{fig:grover_circ}
     \end{subfigure}\\
     \begin{subfigure}[b]{0.23\textwidth}
         \centering
         \includegraphics[width=\textwidth, trim={0cm 0.4cm 0cm 0cm},clip]{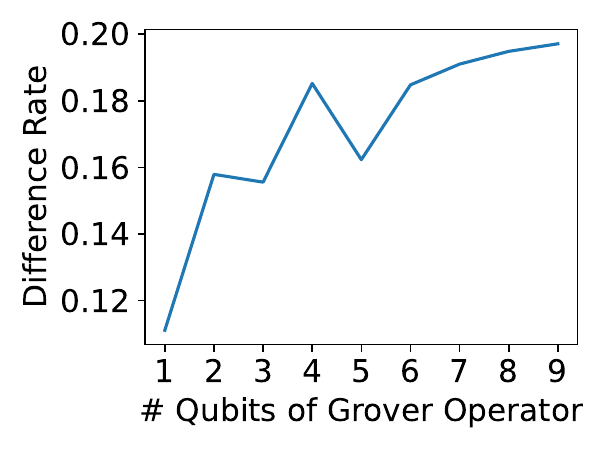}
         \caption{\small The rate of the number of different gates to the number of total gates in Grover operator}
         \label{fig:gate_rate}
     \end{subfigure}
     ~
     \begin{subfigure}[b]{0.23\textwidth}
         \centering
         \includegraphics[width=\textwidth, trim={0cm 0.4cm 0cm 0cm},clip]{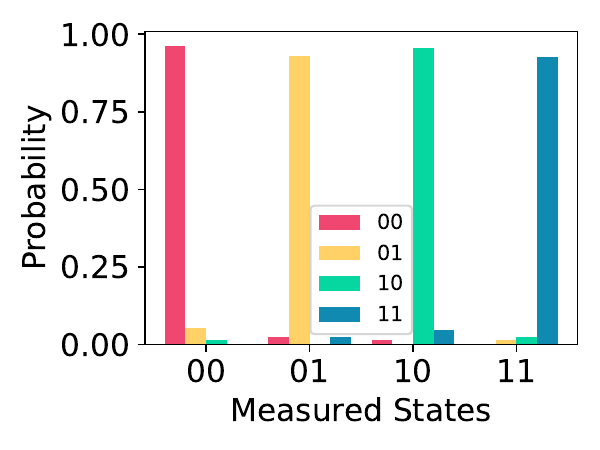}
         \caption{\small Results of 2-qubit Grover's search with targeted states modified by phase mismatch.}
         \label{fig:grover_result}
     \end{subfigure}
    \caption{\small Demonstration on 2-qubit Grover's search.}
    \label{fig:grover}
\end{figure}

This demonstrates the pulse attack to manipulate results.

In the experiment, we implement 2-qubit Grover's search~\cite{10.1145/237814.237866} as the parametric quantum circuits. The whole Grover operator is provided as one custom gate. Grover's search is a quantum algorithm for unstructured search that finds with high probability the unique input to a black box function that produces a particular output value using just $O(\sqrt N)$ evaluations of the function. The parameters we used are the quantum state to search in the computational basis, i.e., $\ket{00}, \ket{01}, \ket{10}, \ket{11}$, and we assume that attackers provide this parametric Grover operator. 

For different search states, the Grover operator is different, as shown in Figure~\ref{fig:grover_circ}. However, only several gates are different, and all of them are rotation-Z gates. The percentages of different gates for search states in the computational basis are shown in Figure~\ref{fig:gate_rate}. It shows that only a small portion of the quantum circuit is different, and the percentage is even smaller when counting pulses rather than gates. Therefore, attackers can only modify a small portion of the Grover operator by phase mismatch to change the search states. The results are shown in Figure~\ref{fig:grover_result}. Inside the whole custom gate for the Grover operator, only the phases of 4 gates (8 underlying pulses) are modified, and the output is successfully redirected to the desired attack states.

\subsubsection{Level 4 Victim: Pulse Attack on Quantum Neural Networks}

\begin{figure*}[ht]
     \begin{subfigure}[b]{0.7\textwidth}
         \centering
         \includegraphics[width=\textwidth, trim={0cm 10.5cm 0cm 0cm},clip]{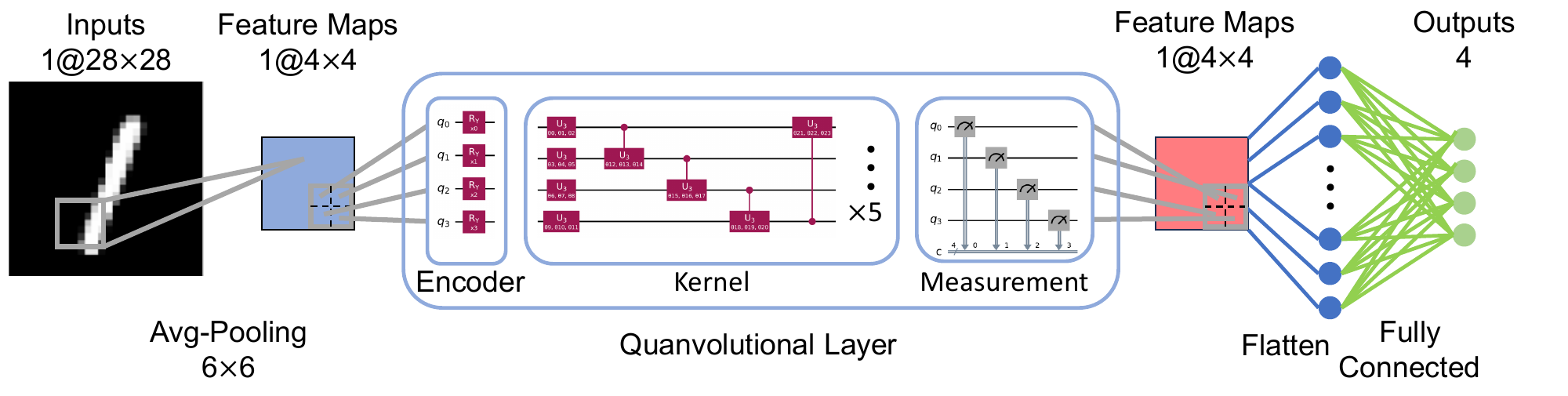}
         \caption{\small Quantum neural network architecture.}
             \label{fig:qnn_arch}
     \end{subfigure}
     ~
     \begin{subfigure}[b]{0.3\textwidth}
         \centering
         \includegraphics[width=\textwidth, trim={0.4cm 0.4cm 0.2cm 0.4cm},clip]{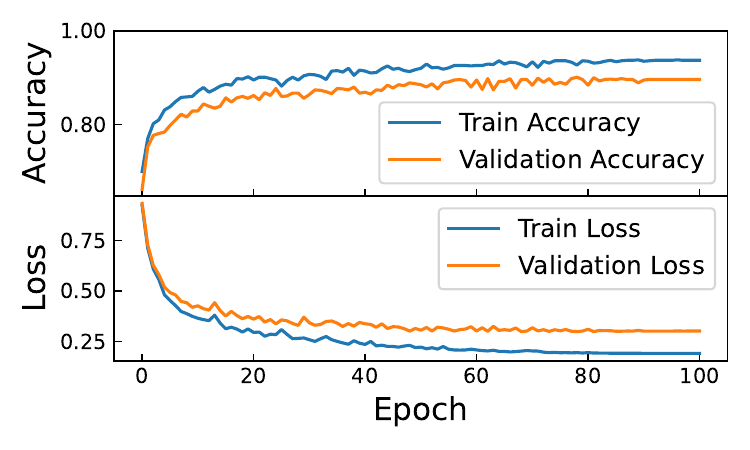}
         \caption{\small Accuracy and loss in training.}
         \label{fig:original_train_acc_loss}
     \end{subfigure}
    \caption{\small The quanvolutional neural network (QNN) architecture and training data in the experiment. The quanvolutional layer is computed on quantum computers, while others are the same and computed in classical computers.}
    \label{fig:qnn}
\end{figure*}

Quantum algorithms are also often implemented with parameters to control the quantum circuits, such as the rotational angles for quantum gates. The parametric quantum circuits are widely used in quantum algorithms that require training or variational operations like quantum machine learning and quantum approximate optimization algorithm~\cite{9951187, liang2024napa, liang2023advantages, liang2023spacepulse}. The parameters can also be parameters for pulses, as they can be directly learned and the quantum circuits can be directly optimized at the pulse level. Similar to classical neural networks, quantum neural networks include many trainable weights, and the final trained weights are not necessarily explainable. In the following, we implement one quantum neural network for image classification and evaluate the pulse attack on it.

We implement the quanvolutional neural networks proposed in ~\cite{henderson2019quanvolutional}, which is similar to the convolutional neural networks (CNN) in classical computing~\cite{6795724}. More specifically, we implement the example quanvolutional neural network provided by TorchQuantum~\cite{hanruiwang2022quantumnas}. The architecture is shown in Figure~\ref{fig:qnn_arch}. We use MNIST~\cite{6296535} as the dataset, which is a database of handwritten digits from 0 to 9. Each image has 28$\times$28 pixels. The quanvolutional layer is a 4-qubit circuit that includes 3 parts: (1) Encoder: one rotational Y gate on each qubit where each rotation angle is one image pixel value. This is used to encode the classical data into quantum circuits. (2) Kernel: one U3 gate on each qubit and one control U3 gate on each adjacent qubit pair. Each U3 gate includes the 3 Euler angles to rotate the qubit, and the control U3 gate is the controlled version of the U3 gate. All 3 angles in each U3 and control U3 gate are trainable parameters to be learned during the training process. The kernel is applied 5 times in a row. (3) Measurement: the quantum circuit results are measured, and the expected values on each qubit will be the output of the quanvolutional layer. Other parts are the same as classical neural networks, including the average pooling layer before the quanvolutional layer and the flatten and fully connected layers after it. Notice that for the training efficiency of QNN, many operations are done to decrease the computational complexity, including the large 6$\times$6 average pooling layer, images with only 4 labels from 0 to 3, and in total 1000 images are used for training and validation. The experiments are done in the simulators. The training process is shown in Figure~\ref{fig:original_train_acc_loss}. With 100 epochs, the model reaches an accuracy of 93.7\% on the training dataset and 89.6\% on the validation dataset.

\begin{figure}[t]
     \centering
     \includegraphics[width=0.50\textwidth, trim={0cm 0cm 0cm 0cm},clip]{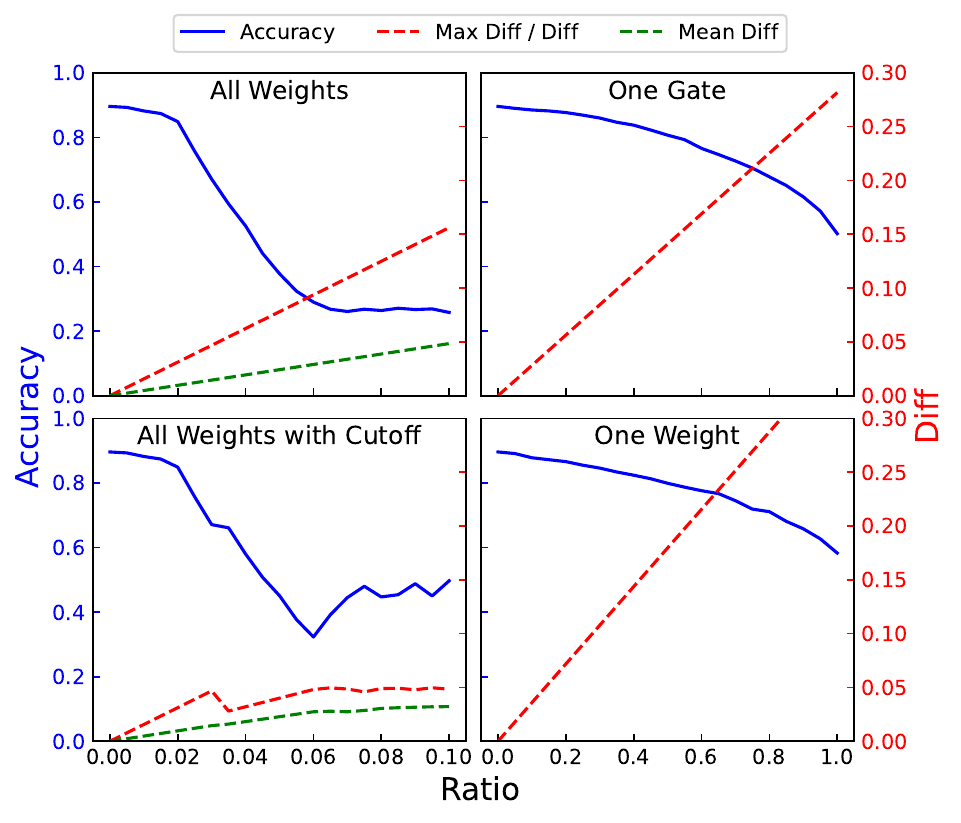}
     \caption{\small The accuracy drops with weight changes in the quanvolutional layer. Diff is the norm difference. The weights are added by the ratio multiplied by their gradients, except for ``One Weight" where the weight is directed added by ratio. ``All Weights" changes weights in all gates. ``All Weights with Cutoff" changes weights in all gates when the norm difference is smaller than 0.05. ``One Gate" changes only the weights in the first U3 gate. ``One Weight" changes the first weight in the first U3 gate.}
    \label{fig:weight_change}
\end{figure}

First, the influence of weight changes on the quanvolutional layer on the accuracy, and the corresponding norm differences, are experimented with. The norm difference is computed as: view 3 parameters in a U3 gate as a 3-dimensional vector, compute the norm of the change divided by the norm of the original vector. We evaluate four ways to change weights: (1) All Weights: all weights are added by the ratio multiplied by their gradients; (2) All Weights with Cutoff: if one U3 gate norm difference is larger than a threshold, then its weights will not be changed. In the experiment, the threshold is set to be 0.05. (3) One Gate: only the first U3 gate is changed. (4) One Weight: only the first rotation angle of the first layer is changed, and the ratio in the figure is the absolute value of the change. The gradients are computed by forwarding only the first epoch of the training dataset.

The results are shown in Figure~\ref{fig:weight_change}, which demonstrates that with a small change, the model results can be influenced much. With a maximum of around 10\% difference and a mean of 3\% difference on all parameters, the accuracy drops from 89.6\% to 26.8\%. Moreover, with the 5\% cutoff threshold to limit the maximum difference, only a mean of 2.7\% change, the lowest point in the figure shows that the accuracy can be decreased to 32.3\%. If changing most weights of the layers is difficult, the accuracy decrease to 60\% can also be realized with around 25\% change for the first rotational gate or 35\% change for only one weight in the first layer. Note that we did not do the experiments to find the best weight to change, but arbitrarily chose the first layer. With careful explorations, a smaller change in the weight may be found and thus the attack can be more covert.

\begin{figure}[ht]
     \centering
     \includegraphics[width=0.45\textwidth, trim={0cm 0cm 0cm 0cm},clip]{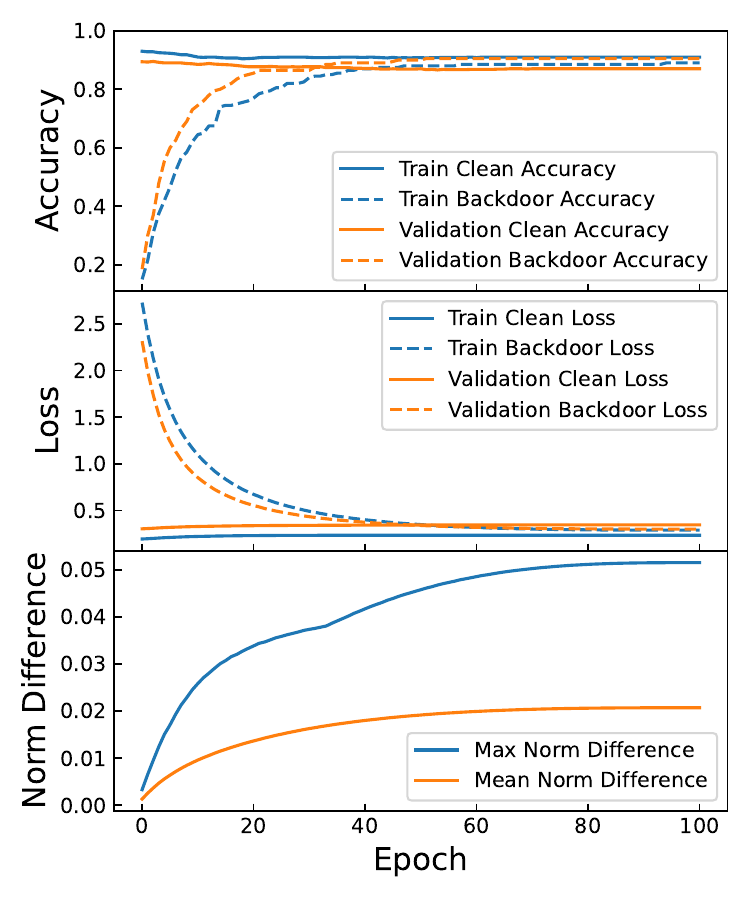}
     \caption{\small The training accuracy, loss, and norm difference in the fine-tuning process of the backdoor attack.}
    \label{fig:backdoor_train_acc_loss}
\end{figure}

More secret attacks can be implemented. Rather than directly changing the weight and destroying the model, some attacks such as the backdoor attack~\cite{gu2019badnets, liu2020reflection} can insert trojan into the model, leading to malicious results only when trojan is triggered. We implement a simple backdoor attack on the previously trained model. We mixed 20\% backdoor samples into the training dataset and fine-tuned the model. The trojan pattern is a 3$\times$3 white pattern on the upper left of the figure. The trojan can be other patterns. This pattern is arbitrarily chosen, and it can be smaller if the average pooling layer is smaller or removed. Images with the pattern are arbitrarily assigned to target 0.

Figure~\ref{fig:backdoor_train_acc_loss} shows the fine-tuning results. Without fine-tuning, the model has an accuracy of 21.6\% on the backdoor validation data. When training around 40 epochs, the model is well-backdoored. The accuracy of the clean training data and validation data is 90.9\% and 87.1\%, with 2.8\% and a 2.5\% drop from the original model, while the accuracy of the backdoor training data and validation data is 87.0\% and 89.0\%, achieving similar accuracy as the clean data. The changes to realize the backdoor attack are small, with a maximum difference of 4.1\% and a mean difference of 1.8\%. The figure shows that further training will not change the weights and accuracy much, i.e., with around 2\% change, the model can be inserted with a trojan.

\section{Defense Framework}
\label{sec:defense}

\begin{figure*}[ht]
     \centering
     \includegraphics[width=0.98\textwidth, trim={0cm 1.35cm 4cm 0cm},clip]{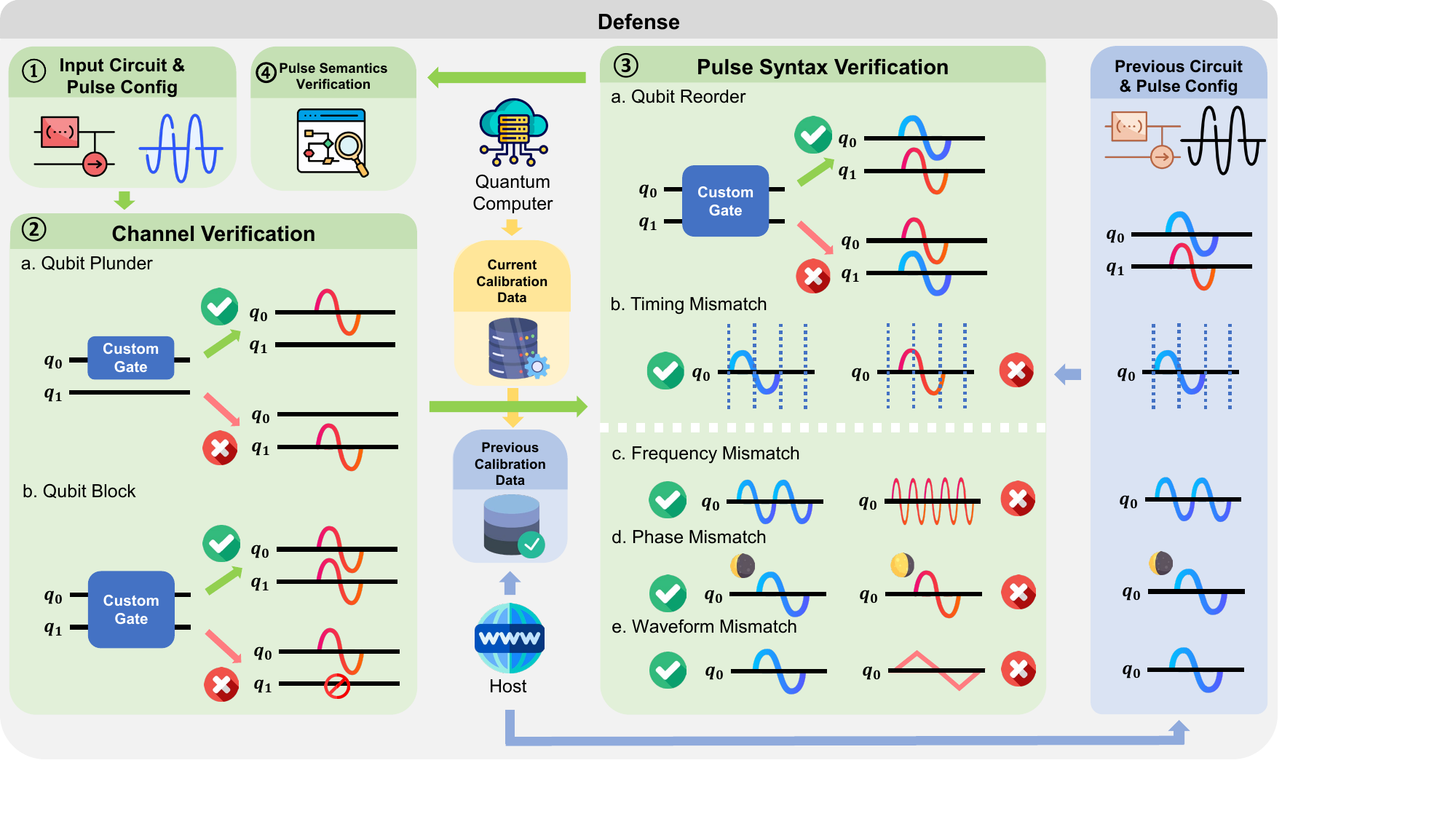}
     \caption{\small Illustration of the defense. With the quantum circuit (and pulse configuration included) as the input, the channel verification can detect the qubit plunder and the qubit block attacks by checking the exact match of the gate-level qubits and the pulse-level channels. For the qubit reorder and the timing mismatch attack, we propose that the circuits must be associated with the previous pulse configuration for instructions comparison. For the remaining attacks, the previous and current calibration data are also needed for parameter comparison.}
    \label{fig:defense}
\end{figure*}

In this section, we propose one framework to defend from the attacks. The verification of quantum programs can be inspired by the verification of classical programs, though with some peculiarities. Many traditional methods can be used and may be helpful to some degree. For example, the quantum program can be checked and distributed with associated hash codes such as SHA256. This is a common way of distributing classical software. This could be helpful when the properties of quantum computers do not have a big change, similar to classical computing. However, as we explained in Section~\ref{sec:challenges_of_syntax_verification}, the volatility of quantum computers may make this method fruitless because the physical properties of the quantum hardware may undergo big changes, and thus the same pulse configuration may not have the same functionality. On the other hand, instead of resorting to the provided pulse data, users may calibrate by themselves to get the up-to-date pulse data. Nevertheless, the cost issue described in Section~\ref{sec:challenges_of_syntax_verification} may discourage users from doing this, and users may also need to acquire the programs for calibration from attackers and these programs may also be tampered with by attackers. 

In fact, not all of the proposed attacks are due to the same reasons. The cause for the attacks can be divided into:

\begin{itemize}[leftmargin=*]
    \item \textbf{Software flaw: } This is the cause of the qubit plunder and the qubit block. These two attacks arise from the fact that current quantum software development kits overlook the correspondence between gate-level qubits and pulse-level channels. As a result, they can be detected and fixed at the software level.

    \item \textbf{Interface defect: } This is the cause of the qubit reorder. The current software design for adding pulse-level controls is deficient as we explained before. The custom gate is just a container used to store the pulse-level controls. Even though we can enforce the correspondence of gate-level qubits and pulse-level channels, we still cannot guarantee the functionality of pulses inside. This attack must be detected if the function of the custom gate is specified beforehand. The loose interface design in most current SDKs may be due to the consideration of convenience in development. The custom gate with the abstract functionality can be applied to any qubits, while the underlying pulse configuration is qubit-dependent. In some sense, they are contradictory and cannot be easily automatically modified, otherwise, the pulse configuration can also be easily updated since the automatic modification requires the pulse syntax verification. The straightforward way to implement the custom gate with the same functionality on all qubits will either require the current design where the custom gate is flexible and pulses are selected for the specific qubits, or the custom gate is inflexible and qubit-dependent. Both of them entail inflexibility and complexity in software development.

    \item \textbf{Challenges for the pulse level verification: } This is the cause of other attacks. Again, the pulse configuration is only a temporary representation of the functionality of the custom gate. The challenges are discussed in Section~\ref{sec:attack_stealthiness}.
    
\end{itemize}

\noindent In the remainder of this section, we provide one straightforward framework for detecting the attacks through \textit{channel verification}, \textit{pulse syntax verification}, and \textit{pulse semantics verification}. Figure~\ref{fig:defense} shows the workflow.

\subsection{Channel Verification}

The qubit plunder and the qubit block can be directly detected with the software given a quantum circuit. For each custom gate in the quantum circuit, the qubits to which the custom gate is applied are provided, and the pulse configuration for the custom gate is also included in the quantum circuit. Therefore, with the mapping between gate-level qubits and pulse-level channels, the qubits and the channels must be the exact match, i.e., there must be at least one pulse on each qubit of the custom gate, and none of the pulses can be on any qubit that is not in the custom gate. Notice that qubit reorder cannot be directly detected with this verification, since the match may already be exact.

Besides, there are many defects in most current SDKs that hinder channel verification. For instance, the register and memory slots for storing the classical information such as the measurement results must also correspond to the custom gate to address the custom measurement. The current SDKs have not counted for this. For instance, the \texttt{Gate} class design in Qiskit does not include the classical bits. In addition, all types of channels must be counted, including drive channels for single-qubit operations, control channels for multi-qubit gates, and acquire and measure channels for measurement. However, this leads to design complexity. For instance, if one custom gate is a 3-qubit gate, it can be one single-qubit gate on each qubit, and thus control channels are blocked and not used, or they can be maliciously plundered; or it can be one single-qubit gate and one two-qubit gate, and thus only one control channel is used. Both cases are valid, but it cannot be determined which case is correct with only the input quantum circuit.

\subsection{Pulse Syntax Verification}

The other attacks cannot be directly detected with only the quantum circuit as the input. As explained, in order to detect these attacks, the syntax and semantics of each pulse in the custom gate must be specified. This is reasonable for custom gates with clear goals, such as the pulse-optimized Toffoli gate. Nevertheless, it is not possible for many other gates, such as the end-to-end quantum neural network. 

The pulse syntax verification is to verify that each pulse in the input pulse configuration functions correctly as desired. The pulse syntax verification may be the most difficult part of the verification for quantum programs. It is a complicated problem that if two calibration data are given, how to verify if these two pulse configurations have similar functionality. The related problem is quantum tomography~\cite{tilly2022variational}: given a black box (the custom gate), how to figure out its functionality? One reason is that there are too many intertwined parameters to specify one pulse. For example, if the frequency of the pulse is the calibrated qubit frequency, then the new frequency may also need to be the new calibrated qubit frequency. However, maybe the original data contains errors, and thus all parameters were measured with bias. In this case, the verification will become very difficult. Similarly, different pulse waveforms may result in the same effect, this is again hard to deal with.

We propose to associate the quantum circuit with both the pulse configuration and the calibration data at the time the pulse configuration is calibrated, or identically the date on which the pulse configuration is calibrated. This associated data should be stored at some trusted authorities. Before users execute the quantum circuit or before the circuit publisher updates the pulse configuration, the input pulse configuration will be compared with the trusted pulse configuration, together with the analysis of the current calibration data from the quantum computer vendors or user custom experiments with the trusted calibration data.

For the pulse configurations that only include the native pulses provided by the quantum computer vendors, the verification can be the exact checking. However, in most cases, the parameters of pulses are specified. Since the pulse data is analog and the calibration process is noisy and erroneous, it is difficult to provide exactly the same functionality check. The tolerance may need to be provided for the verification.

\subsection{Pulse Semantics Verification}

Once the pulse syntax is verified, the semantics verification may be straightforward. Gate-level circuits can be described as the directed acyclic graph (DAG)~\cite{Iten_2022}, and pulse-level circuits may also be described using the same language. Qubits in the gate-level DAGs correspond to channels in the pulse-level DAGs, and the verification problem can be reduced to the graph problems. 

Besides, once the syntax of each pulse is determined, the analog operations can be described similarly to the ``instructions" in classical computing, and thus many verification methods for classical programs may be easily applied to quantum programs.

\subsection{Evaluation}

The defense framework was tested against all the attacks described in Section~\ref{sec:demonstration}, and successfully detected all of them. However, the current implementation of the defense operates using an exact matching approach, allowing for variations in pulse amplitude within a specified tolerance range, as mentioned earlier. This design results in low precision but high recall for detecting malicious modifications: it effectively detects circuits where pulses or even gates have been tampered with, but it struggles to handle circuits that vary but generate identical results as the original circuits. For instance, a pulse intended for an X gate could be split into two pulses for an SX gate, and the current implementation would incorrectly flag this as a malicious circuit.

The trade-off between precision and recall can be adjusted a bit by modifying the tolerance range — higher tolerance increases precision while lowering recall. Usually, as circuits only need to be calibrated by simply adjusting the amplitude and the frequency to follow the current conditions of the environment, this exact matching approach proves useful. Nevertheless, a more advanced implementation that can accommodate variations in other pulse specifications is worth exploring and remains an area for future research.

\section{Discussion and Future Work}
\label{sec:discussion}

One impactful future work is to improve the verification. While pulse semantics verification may resemble the verification for classical programs, pulse syntax verification raises a big challenge. One way to define the syntax may be to return to the matrix representation since all quantum gates should be unitary. Then the pulse check can be computing the similarity between two unitary matrices. However, again, this problem is similar to tomography, where we must find a way to efficiently compute the unitary matrix given the~pulses.

Another important topic is how errors are correlated with tolerance in the defense. This can depend on the size of the custom gate, the type of the parameter, etc. The current defense requires input tolerance, where the relation between it and the results is vague. If users can give the largest error threshold and then the tolerance can be computed, the defense will be more useful and powerful.

In addition, if the pulse parameters can be automatically updated given the calibration data from the cloud provider, then the attacks can be mitigated to a great extent. Superficially, if giving the pulses and checking their functionality is like the P problem, then giving the functionality and providing the pulses is like the NP problem. However, if we fix all pulses and only change some of the parameters, this problem may be greatly simplified. 

Lastly, if we think from the other direction from the qubit properties to the pulse data, because pulse data needs to be updated when qubit properties change, maliciously changing the qubit properties can be a new type of attack. This can be one type of supply chain attack that happens in the manufacturing process, or by directly altering the environment in the data center of the cloud platforms.

\section{Conclusion}
\label{sec:conclusion}
This work presented the first thorough exploration of the attacks on the interface between the gate-level and pulse-level quantum circuits and pulse-level quantum circuits. The attacks presented in this work leverage the mismatch between the gate-level definition and description of the custom gate, and the actual, low-level pulse implementation of this gate. By manipulating the pulse definition numerous attacks are possible as this work proposed: qubit plunder, qubit block, qubit reorder, timing mismatch, frequency mismatch, phase mismatch, and waveform mismatch. The attacks are in part possible because there is a lack of sufficient verification in the current quantum software development kits. This work thus also proposed a defense framework to protect from these attacks. In summary, this work provides insight into the future development of secure quantum software development kits and quantum computer systems.

\ifCLASSOPTIONcompsoc
  \section*{Acknowledgments}
\else
  \section*{Acknowledgment}
\fi
The authors would like to thank IBM and Yale University for providing access to IBM's superconducting devices. This work was also supported in part by NSF grants \nsf{2312754} and \nsf{2245344}. The authors would like to thank the shepherd and the anonymous reviewers for their valuable feedback and comments used to improve the final version of this~paper.

\bibliographystyle{IEEEtran}
\bibliography{bibliography.bib}

\balance
\newpage % The Meta-Review should at least start on a new column

\appendices % if not used earlier

\section{Meta-Review}

The following meta-review was prepared by the program committee for the 2025
IEEE Symposium on Security and Privacy (S\&P) as part of the review process as
detailed in the call for papers.

\subsection{Summary}

This paper explores implementation-level issues in the emerging world of quantum computing. Modern quantum computers implement quantum gates as microwave pulses, the exact specification of which vary according to calibration of different machines and over time. The authors show that, by precisely corrupting these pulse width specifications, attackers can make certain malicious changes to quantum circuits, without modifying the gate-level specification of the quantum circuit.

\subsection{Scientific Contributions}
\begin{itemize}
\item Independent confirmation of important results with limited prior research.
\item Identifies an impactful vulnerability.
\item Establishes a new research direction.
\end{itemize}

\subsection{Reasons for Acceptance}
\begin{enumerate}
\item The paper is the first to propose this set of attacks, making it an interesting link in the chain stretching back to Ken Thompson's 1984 observations on Trusting Trust.
\item The authors validated the effectiveness of attacks both via simulation and with devices, making it clear that the attacks are feasible.
\item The paper includes a good description of critical background that many readers may lack, despite the complexity of the subject matter.
\end{enumerate}

\end{document}